\documentstyle[aps,prl,multicol]{revtex}

\begin{document}

\title{ Fermi Surface Topology of $Bi_{2}Sr_{2}CaCu_{2}O_{8+\delta}$
at $h\nu=33eV$: hole or electron-like?}

\author{A. D. Gromko$^1$, Y. -D. Chuang$^1$, D. S. Dessau$^1$, K. 
Nakamura$^{2}$, Yoichi Ando$^{2}$}
\address{$^1$Department of Physics, University of Colorado, 
Boulder,Colorado, 80309-0390}
\address{$^2$Central Research Institute of Electric Power Industry
(CRIEPI), 2-11-1 Iwato-Kita, Komae, Tokyo 201-8511, Japan}

\maketitle


\begin{abstract}
We present new results from Angle-Resolved Photoemission experiments
(ARPES) on overdoped $Bi_{2}Sr_{2}CaCu_{2}O_{8+\delta}$ (BSCCO) crystals. With
greatly improved energy and
momentum resolution, we clearly identify the existence of
electron-like portions of Fermi Surface (FS) near $\bar M$ at $h\nu=33eV$.
This is consistent with  previously reported data and is robust
against various FS crossing criteria.  It is not due to an artifact
induced from $\vec k$-dependent matrix element effects.
We also present evidence for a breakage in the FS pointing
to the possible existence of two types of electronic components.

\end{abstract}
\vspace*{-0.3 in}
\begin{multicols}{2}
\narrowtext

Angle-Resolved Photoemission Spectroscopy (ARPES) has become one of
the most powerful tools for understanding the physics and electronic
structure of high temperature superconductors (HTSC) and other
correlated electron systems since it allows one to probe the energy and
momentum relations directly. Over the past decade, major discoveries
have been made on both the normal and superconducting states which
include the Fermi Surface topology, superconducting gap symmetry,
normal state pseudogap, etc. \cite{1}.

Among the normal state properties, the Fermi Surface (FS) topology is
one of the most important since it needs to be determined prior to
correctly predicting many physical properties. Most of our
information about the FS topology of HTSC's has come from ARPES
studies on BSCCO, and the results have been widely interpreted as a
hole-like barrel centered around the $(\pi,\pi)$ or X(Y) points of
the Brillouin zone, as illustrated in figure 1(a) \cite{1,2,2.5,3}. 
This conclusion was made
mainly by using incident photon energies around $21 eV$ \cite{2,2.5}.
Recently, we showed that the spectra and the
physical picture appear quite different when measured using
$33 eV$ photons- there is  a strong
depletion of spectral weight around $\bar M$ $(\pi,0)$ and the
FS appears to have electron-like portions, as illustrated in
figure 1(b)\cite{4}.  This result
was later confirmed by Feng et al. \cite{5}.

This new interpretation of the data was questioned by two
experimental groups in four recent papers \cite{6,7,8}.
Fretwell et al. presented a detailed two-dimensional FS mapping on
optimally doped BSCCO using $33 eV$ photons\cite{6}. Although their data
reproduced the salient features found in~\cite{4,5},
they attempted to explain the data under the framework
of the conventional hole-like FS centered on $(\pi,\pi)$ by
invoking empirically determined $\vec k$-dependent matrix element
effects. They further argued that the poor energy and momentum
resolutions combined with limited sampling of the Brillouin Zone in
previous experiments led to
incorrect interpretations.  A similar interpretation was
given by Mesot et al.\cite{7} on overdoped $Bi_{2}Sr_{2}CuO_{6+\delta}$
using symmetrization arguments on lower-resolution data. They
stated ``the FS is hole-like and is independent of photon energy.''

In order to clarify these issues, we have carefully probed the $\vec 
k$-space region
around $\bar M$ on overdoped $Bi_{2}Sr_{2}CaCu_{2}O_{8+\delta}$ at
$h\nu=33eV$ with very high energy and momentum resolution.  Our raw data is
similar to that presented by Fretwell et al., although ours was taken at
$100K$ in the normal state - a more ideal experimental situation
(compared to their superconducting state data) for probing FS 
topology.  We analyze
the data in a number of ways, all of which give results consistent with the
electron-like topology originally proposed by us \cite{4}.  In contrast the
hole-like topology presents a much poorer fit to the data.  We also show that 
Fretwell et
al.'s data is consistent with the
electron-like FS topology, further supporting our arguments.  We also
have performed the symmetrization method to determine FS crossings,
and in contrast to Mesot et al.'s data \cite{7}, our symmetrized
data does support the electron-like topology.  In this sense our data
is different from theirs, possibly due to uncertainties in their Fermi
energy calibration.

Borisenko et al. and Golden et al. \cite{8} recently presented beautiful and
highly detailed two-dimensional FS mappings on optimally doped BSCCO using
$21.22 eV$ photons.
While their data gives perhaps the clearest evidence yet for the 
hole-like topology in this
photon energy range, it does not address the different behavior
observed at $33 eV$.  Golden et al. did show a small portion of data at
$33 eV$ consistent with the data of Chuang et al. \cite{4}, and
similar to Fretwell et al. and Mesot et al. \cite{6,7} they indirectly 
argued that this
data was affected by unfavorable matrix element effects.

The experiments were done at Beamline 10.0.1 at the Advanced Light
Source (ALS), Berkeley, CA using a Scienta SES 200
energy analyzer.  We used the angle mode of the analyzer to 
simultaneously collect
89 individual spectra along $14^{o}$ wide angular slices.  We present 11 of
these slices, representing almost 1000 individual
energy distribution curves (EDCs) from one sample. The
angular resolution along these slices was about $\pm 0.08^{o}$ in
the $\theta$ direction and
about $\pm 0.25^{o}$ in the perpendicular $\phi$ direction.  At 
$h\nu=33eV$, the
converted momentum resolution is $(k_{x}, k_{y}) \approx
(0.01\pi, 0.03\pi)$.  We could map out the two-dimensional Brillouin 
zone by rotating
the sample in either $\theta$ (parallel to the $14^{o}$ slice) or in
$\phi$ (perpendicular to the slice).  The analyzer was left fixed with
the central analyzer angle making an $83^{o}$ angle of incidence relative to
the photon beam. In this configuration, the $14^{o}$ slices are
parallel to the incident photon polarization direction and the 
$\Gamma-\bar M$ high
symmetry line.

The energy resolution was better than
$10meV$ FWHM, as determined by the 10-90$\%$ width of a gold reference
spectrum taken at $10K$.  The sample Bi046 used in this study was 
about 3mm on a side and was annealed
in oxygen to overdope it, giving a $T_{c}=79K$.
Throughout the whole experiment, the base pressure was maintained
below $4*10^{-11}$ torr and the temperature was at $100K$, well above
$T_{c}$.  All experimental data were normalized by using the high-harmonic
emission above $E_{F}$, as discussed in reference \cite{4}.

Figure 2 shows the new data from this sample.  The 11 panels
of part (a) show false-color Energy Distribution Curves (EDC) taken
at 2 different sets of $\theta$ angles (left and right panels) and 9 different
$\phi$ angles. The vertical axes are the binding energy, and the
horizontal axes are the $\theta$ angle along the $14^{o}$
slice, with $0^{o}$ equal to normal emission.  Each of the plots was
normalized separately, i.e. the color scale can not be connected from one
plot to another.

Each of these plots shows one or more features which disperse in
energy as a function of angle (or
wavevector $\vec k$).  Most of these features can
be easily followed up to $E_{F}$ at which
point the features disappear due to the FS crossing.  As a guide to
the eye, we have overlayed a black curve on top of the data and 
labeled
each features as S.S. (superstructure band) or M (main band). Due
to polarization selection rules \cite{9}, the ARPES features along 
the $\Gamma-Y$
line have unfavorable emission such that in panels (ix)-(xi) the
S.S. band has stronger intensity than the main band. In figures
2(b) and 2(c) we plot white dots which indicate the FS crossing 
points determined
by looking at the dispersion in part (a), on the $\Gamma-\bar M-Y$ 
quadrant of the
Brillouin zone.  (The position of the 11 individual slices are indicated
in panel (b).)

Another way to visualize FS crossings is to make a two-dimensional
plot of the spectral intensity at $E_{F}$, i.e. $A(\vec k,E_{F})$. 
This is plotted
in parts
(b) and (c) with data obtained by compressing the 11 panels of part (a).
For these plots the relative intensity between each of the 11 panels is
critical. They were first normalized by looking at the high-harmonic
emission above $E_{F}$ only. We then integrated the spectral
weight of the EDCs of Figure 2(a) over a $50meV$ energy window
centered at $E_{F}$.  Figures 2(b) and 2(c) show false color
plots of this spectral intensity as a function of $\theta$ and $\phi$
with the color scale on the left.
The FS should show up on this plot as the region of maximum spectral
intensity.

The black lines in figure 2(b) show the experimentally determined FS
from this data.  The thick black lines represent the main FS in the
first and second zones, while the thin lines represent the
superstructure-derived FS, which is obtained by shifting the main FS
by $(0.2\pi, 0.2\pi)$ along the $\Gamma-Y$ direction.  These FS's
are consistent with the white dots obtained from
panel (a) as well as with the high intensity locus of panel b.
They also are consistent with the symmetrization method (figure 3(c)) and
with the gradient or 50$\%$ point of $n(\vec k)$
plots \cite{4,5}. The FS determined here is qualitatively and
quantitatively (to better than 5$\%$) the same as that determined in reference
\cite{4}. The very slight shift of the high intensity locus away from
$\bar{M}$ compared to the white dots is due to an
effect of the finite energy integration window used to make the plot
of panel (b) \cite{10}.

In contrast, the hole-like FS topology overlaid on the data in panel (c)
cannot explain many of the crossing
locations.  This FS is taken from Fretwell et al, who also took
data at $33eV$ \cite{6}.  The thick black line is the main FS while 
the thin black and
red lines are the first and second order S.S. FS's. The yellow
lines are possible shadow bands obtained by reflecting the main and
S.S. FS's about the $(\pi,0) - (0,\pi)$ line\cite{2.5,8}. 
These FS's show dashed sections which,
as proposed by Fretwell, indicate a strongly reduced spectral intensity
region due to strongly $\vec k$-dependent matrix
element effects \cite{6}. These are supposed to account for the lack
of observation of a FS crossing from $\bar{M}-Y$.  Even though this 
suggestion has
many more free parameters than the electron-like topology of panel
(b), it does not match the data as well. In particular it cannot match the
curvature of the high intensity portion near $(\theta,\phi) =
(14,2)$ (also see cut (iii) in panel (a))
and it has trouble with the portion of the FS naturally
explained by the S.S. band in panel (b) (the part circled
in white). In the hole-like topology of figure 2(c) this would have to
be explained by a combination of three bands - black,
red, and yellow.  The
unlikeliness of this is amplified when a closer look at the EDC's is
taken. For example, panel (c) shows that the crossing at $(\theta,\phi) =
(19,5)$ should come from the yellow shadow band.  As such it should have
reversed E vs. $\vec k$ dispersion from the main band, while panel
(vi) of part (a) shows that it does not.  The experimental intensity 
is also too
strong - it should be a shadow of a S.S. band and should
also be strongly reduced by matrix element effects (dashed lines).

The $33 eV$ FS plots presented by Fretwell et al. \cite{6} are quite similar
to the data in figure 2, one
of which is reproduced in figure 3.  Panel (a) shows their data and
their interpretation within the hole-like
FS topology. The image plot was obtained by integrating the spectral 
weight over
the relatively large energy window $(-100meV,100meV)$ in the 
superconducting state at $40K$.
We feel that the hole-like topology presented in this figure has a number of
deficiencies and can not explain the data well.  First, the hole-like 
FS segments extend
towards $\bar{M}$ while the data does not. Fretwell et al.
attempted to reconcile this by empirically
introducing a strongly $\vec k$-space dependent matrix element
effect to drastically reduce the weight near $\bar{M}$. However such
an effect cannot explain the curvature of the FS which is manifested
in the intensity plot by the locus of high intensity points. Namely,
in figure 3(a) we have circled a high intensity region which cannot
be explained by a hole-like FS.  Figure 3(b) shows the electron-like
FS (thick white line)
plus S.S. band portions (thin white lines) overlayed with the same 
intensity plot. Now
the main band FS trace matches beautifully with the locus of high
intensity spots, without having to introduce any complicated
matrix-element dependent physics.

The S.S. band circled in our figure 2(b) is not apparent in Fretwell
et al.'s data of figure 3(a) and (b).  We
suggest two possible reasons for this discrepancy: (1) the S.S. bands
may be weaker in Fretwell's data, or (2) they perhaps did
not include enough dynamic range in their color scale plot, so that
the S.S. bands were not apparent.

Using symmetrized EDC's, Mesot et al. \cite{5} also argued that their 
$34eV$ data
from overdoped $Bi_{2}Sr_{2}CuO_{6+\delta}$ supported the hole-like
topology since they did not observe a FS crossing along $\Gamma-\bar M$
but they did observe one along $\bar M-Y$.  We have applied this
method to our data, and find that it is supportive of an
electron-like topology, as shown in figure 3(c).  Each EDC was added to a
mirror image of itself reflected around $E_{F}$, which assuming
electron-hole symmetry near $E_{F}$ will remove the
effect of the Fermi function cutoff.  As stressed by Mesot, the symmetrized EDC
will show a peak at $E_{F}$ if there is a FS crossing, otherwise it will
show a dip.  Figure 3(c) shows the symmetrized EDC along
$\Gamma-\bar{M}-Z$ from the data of figure 2. The dip at $E_{F}$ 
disappears at around
angle $(\theta,\phi)=(15,0)$, indicating the FS crossing - a location
that agrees to within $2\%$ with that obtained
directly from figure 2.  The consistency of the results obtained by 
all methods (see also
references \cite{4} and \cite{5}) gives us confidence in the 
electron-like topology
observed at $h\nu=33eV$.  However, we still need to worry about the
different result obtained by Mesot et al.  The difference might be due
to an incorrect determination of $E_{F}$ since the symmetrization 
method is very
sensitive to a shift of $E_{F}$. We have performed symmetrizations on
BSCCO data from three different experimental systems and from both the
single layer and double-layer compounds. All data from over or
optimally doped samples have given the same result - a peak at
$E_{F}$ after symmetrization, indicating a FS crossing along $\Gamma-\bar M$.
This gives us confidence that an undetected $E_{F}$ calibration error
could not have adversely affected our data.

The increased energy and momentum resolution obtained in figure 2
brings up other subtleties which have not been previously observed.  If we
track the main electron-like FS from the $\Gamma-Y$ line towards the
$\Gamma-\bar M$ line, the intensity first gets greater, which is
understood due to polarization effects \cite{9}.  It then gets weaker 
around $(14,2)$
before getting stronger again along the $\Gamma-\bar M$ line.  This makes the
FS appear as if it has two components - one nearer the $\Gamma-Y$ line
and one nearer the $\Gamma-\bar M$ line.  Further study needs to be
carried out to deconvolve the origin of the separation of the FS into
these components, as well as to study potential differences in the
behavior of each component.

Finally, of course, there is the critical issue of connecting the
electron-like FS topology observed at $33 eV$ with other topologies
observed at other photon energies.  A natural possibility is to
consider coherent three-dimensional band structure effects from the 
$x^{2}-y^{2}$ band (or from the $z^{2}$ band \cite{Jason}), although this
would need to be reconciled with the highly two-dimensional nature of the
cuprates.  Even-odd splitting between the $CuO_{2}$ bilayer may produce
both an electron and a hole like FS in the same sample, although this
would need to be reconciled with the single-layer Bi2201 data which
also appears to show both topologies \cite{4}.  Two FS's may simultaneously
exist in the same sample for other reasons as well, for instance due
to phase separation into hole-rich and hole-poor regions or into
regions with and without stripe disorder \cite{5}, each of
which may produce its own FS portions.  Within these scenarios
we still need to understand why one piece of the FS is accentuated at
one photon energy while another is accentuated at another.  Matrix
element effects may play a role in this \cite{11}.
 
In this communication we have presented high energy and momentum
resolution ARPES results on the normal state of overdoped
$Bi_{2}Sr_{2}CaCu_{2}O_{8+\delta}$  at $h\nu=33eV$.
We clearly identify the existence of electron-like portions of Fermi
Surface near $\bar M$ by looking at the high intensity locus in
$A(\vec k,E_{F})$ plots, dispersion of EDCs, and symmetrized EDCs. We
reach the same result by looking at either the main band or the
S.S. bands. In contrast, the hole-like topology cannot
explain the details of the spectra, even
with the ad-hoc inclusion of strong $\vec k$-dependent matrix element
effects. In addition, our increased resolution shows some new and
potentially important subtleties of the data including a break
in the main FS.

We acknowledge helpful discussions with Z.-X. Shen and Y. Aiura, 
experimental help
from S. Keller, P. Bogdanov and X.-J. Zhou, and analysis software from
J. Denlinger.  This work was partially supported by a grant from the ONR.
The ALS is supported by the DOE.

\end{multicols}
\vspace*{-0.2in}

\begin{figure}
   \caption{Hole-like FS topology (a) versus electron-like
   topology (b).}
\end{figure}
   \vspace*{-0.2in}

\begin{figure}
    \caption{(a)
    E vs. $\vec k$ plots from overdoped Bi2212 sample Bi046 measured at $100K$.
    Vertical axes are binding energy ($eV$)
    and horizontal axes are the angle $\theta$.  The right panels
    are centered at $\theta =18^{o}$, while the left panels 
    are centered at $\theta =8^{o}$.  
    The location of each cut is shown in panel (b).
    The color scale is shown
    at the top left corner, with larger value (1.0) indicating
    more spectral weight. Black lines indicate the E vs. $\vec k$ dispersion
    relation and are guides to the eye. The FS
    crossing points are determined by looking at the intersection of
    the guide lines and $E_{F}$ and are
    labeled S.S. (superstructure band) and M (main band).
    Panels (b) and (c) are plots of the spectral intensity at $E_{F}$.
    White dots are the FS crossings from part (a). Panel
    (b) shows the agreement with the electron-like topology while panel (c)
    shows the disagreement with the hole-like topology.
  }
\end{figure}
   \vspace*{-0.2in}

\begin{figure}
   \caption{(a) Data from Figure 1(c) of Fretwell et al.[6] The
   circled region cannot be explained by the the hole-like FS
   topology. (b) Same data overlayed by a FS with electron-like portions
   (white). (c) Symmetrized EDCs along $\Gamma - \bar M$ for a variety
    of $\theta$ angles, with $\phi$=0.  The first peak at $E_{F}$ shows up at
    $(15,0)$, indicating a FS crossing.   }
\end{figure}
   \vspace*{-0.2in}


\begin{references}
\bibitem{10} Reference \cite{4} shows that a finite energy
integration window makes $A(\vec k,E_{F})$ plots approach $n(\vec k)$
plots.
\end{references}
\end{document}